\documentclass[aps,prd,twocolumn,showpacs,nofootinbib]{revtex4-1}
\usepackage{graphicx}
\usepackage{amsmath}
\usepackage{amssymb}
\usepackage{bm}
\usepackage{bbm}
\usepackage{color}

\begin{document}

\title{A proposal on complementary determination of the effective electro-weak mixing angles via doubly heavy-flavoured hadron production at a super Z-factory}

\author{Xu-Chang Zheng}
\email{zhengxc@cqu.edu.cn}
\affiliation{Key Laboratory of Theoretical Physics, Institute of Theoretical Physics, Chinese Academy of Sciences, Beijing 100190, China.\\
School of Physical Sciences, University of Chinese Academy of Sciences, Beijing 100049, China.\\
Department of Physics, Chongqing University, Chongqing 401331, P.R. China.}
\author{Chao-Hsi Chang}
\email{zhangzx@itp.ac.cn}
\affiliation{Key Laboratory of Theoretical Physics, Institute of Theoretical Physics, Chinese Academy of Sciences, Beijing 100190, China.\\
School of Physical Sciences, University of Chinese Academy of Sciences, Beijing 100049, China.\\
CCAST (World Laboratory), Beijing 100190, China.}
\author{Tai-Fu Feng}
\email{fengtf@hbu.edu.cn}
\affiliation{Department of Physics, Hebei University, Baoding 071002, China.\\
Key Laboratory of Theoretical Physics, Institute of Theoretical Physics, Chinese Academy of Sciences, Beijing 100190, China.}

\begin{abstract}
To test the Standard Model (SM) precisely and to look for the clues deviating from SM, a proposal for determining the effective electro-weak mixing angles ${\rm sin^2\theta_{eff}^f\;}$ (particularly f=lept, c, b) is proposed. It is via observing the asymmetries (the forward-backward one $A_{FB}$, the left-right one $A_{LR}$ and the combined left-right forward-backward one $A^{FB}_{LR}$) of the doubly heavy-flavoured hadrons in the production at a super Z-factory (an $e^+e^-$ collider designed with a luminosity as high as possible for modern-day techniques runs around the center-of-mass energy $\sqrt s=m_Z$). To see the sensitivity, uncertainties as well as the event accumulation precisely, the observables for describing the asymmetries of the produced hadrons are computed numerically with varying effective mixing angles, and based on the result analyses, it is concluded that the proposal may offer an independent complement determination of the flavored effective mixing angles.\\

\end{abstract}


\maketitle

\section{Introduction}

In the Standard Model (SM) the electro-weak mixing angle ($\sin\theta_W$) is a common fundamental parameter,
which describes the mixing of the gauge boson fields $W^0$ and $B^0$ into the photon $\gamma$ and the weak boson $Z$.
The mixing angle in the model, being unique and independent of the flavour, appears in the couplings between boson $Z$ and
the neutral weak currents, which are constructed by up-, down-type and left-, right-handed leptons and quarks in a definite manner, and
it needs to be determined by experiments. Whereas in principle, the electro-weak mixing angle determined by experiments may not be `flavour-independent', i.e. it may depend on the flavours of the fermions which appear in the neutral currents. Therefore experimentally
to determine the `mixing angle' precisely and to see whether it depends on the flavours or not are important and may explore
certain clues beyond SM.

In fact, the experimental determination of the mixing angle is directly to measure the so-called `effective electro-weak mixing angles', ${\rm sin^2\theta_{eff}^f}=\kappa_f{\rm sin^2\theta_W}$, where ${\rm sin^2\theta_W}$ is the theoretical mixing angle of SM, $\kappa_f$ denotes the effects of the high order corrections and possible deviations from SM as well, and $f$ denotes the relevant flavour of the fermion.

Having completed the investigation on the asymmetries of the produced doubly heavy hadron in the inclusive production: $e^+e^-\to B_c(B_c^*)+\cdots$ and $e^+e^-\to H_{(QQ'q)}+\cdots$ at the $Z$-boson resonance£¬i.e. the inclusive production of the doubly heavy hadron (mesons $M_{(Q\bar{Q'})}$: $B_c$, $B_c^*$, etc,; and baryons $H_{(QQ'q)}$: $\Xi_{cc}$, $\Xi_{bc}$, $\Xi_{bb}$, $\Omega_{cc}$, etc.) at a $Z$-factory it is found that the asymmetries are quite sensitive to the `effective flavoured electro-weak mixing angles', although the production cross section is not very large. Namely when the weak-boson $Z$ couples to the neutral currents constructed by the specific $f$-flavour fermion (lepton or quark) with the effective electro-weak mixing angles ${\rm sin^2\theta_{eff}^f}$ in the `manner' as that in SM, the investigation indicates that at $Z$-boson resonance there are also sizable asymmetries (the forward-backward one, the left-right one and the combined left-right forward-backward one) of the produced doubly heavy hadron in the inclusive processes $e^+e^-\to B_c+\cdots$ and $e^+e^-\to H_{(QQ'q)}+\cdots$. In fact, it is an additional one to the known situation where there are similar asymmetries for the produced $f$ in the production $e^+e^-\to `Z' \to f\bar{f}$ (here $f$ represents an $f$-flavour lepton or quark).

Therefore here we would like to propose a proposal for experiments to determine the effective electro-weak mixing angles via measuring the asymmetries of the produced doubly heavy-flavour hadron, especially the determined angles may deviate from that of SM. The proposal particularly highlights the $lept$, $c$ and $b$ flavour dependence of the effective mixing angles, but the proposed determination only at a super $Z$-factory (a $Z$-factory with the highest luminosity under modern techniques) can be done owing to the comparative smallness of the cross sections for the relevant production.

In fact, when LEP-I and SLC, as previous $Z$-factories, run at $Z$ resonace, the effective mixing angles ${\rm sin^2\theta_{eff}^f}$ were determined via measuring the forward-backward asymmetry of the fermion (quark or lepton) $f$ (or th anti-fermion $\bar{f}$) of the production $e^+e^-\to `Z' \to f\bar{f}$ by the experimental collaborations \cite{LEP1,LEP2,LEP3,LEP4,LEP5,LEP6,LEP7,LEP8,SLC4}. Additionally when the polarized colliding beams were available at SLC, the mixing angles were also determined via measuring the left-right asymmetry and the combined left-right forward-backward asymmetry for the $f\bar{f}$ production \cite{SLC4}. The determined results on the flavoured effective mixing angles at LEP-I and SLC are ${\rm sin^2\theta_{eff}^{b}=0.281\pm0.016}$ and ${\rm sin^2\theta_{eff}^{c}=0.2355\pm0.0059}$\cite{Zexp}. The value for ${\rm sin^2\theta_{eff}^e}$ determined from the leptonic asymmetry at LEP-I and that from the left-right asymmetry at SLC are different in 3.2 standard deviations, and the determined value ${\rm sin^2\theta_{eff}^{b}}=0.281\pm0.016$ is larger than the value ${\rm sin^2\theta_{eff}^{b}}=0.23293\pm^{0.00031}_{0.00025}$ which is determined from other experiments \cite{Zexp,PDG}. Later on for convenience, the LEP-I and SLC determinations on the mixing angles via the production $e^+e^-\to `Z' \to f\bar{f}$ are called traditional ones.

The traditional determinations via measuring the asymmetries of the $f$-flavour quark in the production $e^+e^-\to (\gamma/Z) \to f\bar{f}$ ($f=c,b$) need to ensure that the events must be of the two-jet production, the two jets in the production are formed by the QCD evolutions of the ¡®primordial¡¯ quarks $f$ and $\bar{f}$, and the flavours of the primordial quarks $f$ and $\bar{f}$ are recognized by detecting the decay products of the leading particle of the jets, etc. Moreover the direction of the produced jet axis or say the direction of the produced quark $f$ (or $\bar{f}$) needs to be determined precisely, but owing to the limited efficiency on identifying the flavour and determining the direction of the `primordial produced quark' $f$ (or $\bar{f}$) in the production $e^+e^-\to f\bar{f}$ (especially when the direction is close to $\frac{\pi}{2}$), the systematic errors in the traditional determinations by LEP-I and SLC are so substantial that they cannot be suppressed much. Obviously to test SM and to search for the clues beyond SM, fresh and more precise experimental determinations on the mixing angles are strongly desired. Hence we propose a new approach to determining the effective electro-weak mixing angles ${\rm sin^2\theta_{eff}^f}$, that is, to measure the asymmetries of the produced doubly heavy hadron ($B_c$ or $H_{QQ'q}$) in the inclusive production $e^+e^-\to B_c+\cdots$ and $e^+e^-\to H_{QQ'q}+\cdots$ ($Q,Q'=b,c; q=u,d,s$) at a $Z$-factory. For this proposal, the flavour and out-going direction of the produced doubly heavy hadron are observed directly, so in terms of the vertex detector, measuring the decay products of the produced hadron, etc., the main systematic errors from determination of the direction and the flavour of the doubly heavy hadron for the determination are strongly suppressed.

The calculations of the doubly heavy hadron production, $e^+e^-\to B_c+\cdots$ and $e^+e^-\to H_{QQ'q}+\cdots$, at a $Z$-factory can be found in Ref.\cite{doublyhadron,doublyhadron1} so here for the proposal we would not repeat the details but collect the useful formulas and results.

The results of Ref.\cite{doublyhadron,doublyhadron1} show that the differential cross-sections for the doubly heavy hadron production display asymmetries relating to the electro-weak mixing angles, and the new experimental results\cite{LHCb114} indicate that the mechanism adopted in Ref.\cite{doublyhadron,doublyhadron1} for the production works well, thus based on the mechanism we have quantitatively studied the capability and possibility, especially, those about measuring the production, $e^+e^-\to B_c+\cdots$ and $e^+e^-\to H_{(QQ'q)}+\cdots$, at a $Z$-factory in determining the mixing angles ${\rm sin^2\theta_{eff}^f}$. The experiences of LEP-I\cite{LEP1,LEP2,LEP3,LEP4,LEP5,LEP6,LEP7,LEP8} and SLC\cite{SLC4} tell us that the experimental errors and the theoretical uncertainties for determining the electro-weak mixing angles may be cancelled or suppressed greatly when the asymmetry observables $A^f_{FB}$, $A^f_{LR}$ and $A^{f,{FB}}_{LR}$ of the $e^+e^-\to f\bar{f}$ production are adopted in the determinations at the $Z$-factories LEP-I and SLC, so for the proposal to suppress errors and uncertainties we define and adopt the observables, $A_{FB}$, $A_{LR}$ and $A^{FB}_{LR}$ which are similar to those adopted at LEP and SLC, to evaluate the asymmetries of the produced doubly heavy flavour hadron, $B_c$ meson or baryon $H_{(QQ'q)}$, in $e^+e^-\to B_c+\cdots$ or $e^+e^-\to H_{(QQ'q)}+\cdots$ ($Q=c,b; Q'=c,b; q=u,d,s$) respectively. Finally based on the analysis on the numerical results of the studies, we conclude that the proposal may offer an independent determination of the effective electro-weak mixing angles and can complement the LEP-I and SLC traditional ones, especially the dependence of the angles on the flavour of $c$ or $b$.

Since the cross-sections of the concerned production are not very large, to collect numerous events of the production so as to increase the statistic precision is crucial for the proposal. Thus the luminosity of the $Z$-factory where the determination is carried out should be as high as possible, i.e. the proposal must be applied at a super $Z$-factory, that means at LEP-I or SLC (the previous $Z$-factory) the proposal cannot be applied due to their low luminosity. Generally, the luminosity of a super $Z$-factory, such as FCC-ee, CEPC and ILC, with modern techniques can be so high as $L\geq 10^{35}cm^{-2}s^{-1}$.  The rate of collecting the so-called `golden events' for determining the effective angles at such a super $Z$-factory is estimated, and based on the rate the conclusion is that at a super $Z$-factory the statistics precision for the proposal may reach a quite good level. Moreover, the recent observation of $\Xi_{cc}^{++}$, as the first one of the doubly heavy baryons, by LHCb \cite{LHCb17}\footnote{The doubly heavy baryon $\Xi_{cc}^+$ was reported to be observed by SELEX collaboration several years ago\cite{selex1,selex2,selex3}, but the SELEX observation was not confirmed later on.} makes the proposal more confident.

The paper is organised as follows: following the Introduction, in Section II we present the useful formulas. On the production of the doubly heavy meson $B_c$ at a $Z$-factory, formulas are merely based on the factorization theorem of NRQCD\cite{nrqcd}; whereas on the production of the doubly heavy baryons $H_{(QQ'q)}$ at a $Z$-factory, not only the factorization theorem of NRQCD\cite{nrqcd} for the doubly heavy di-quark core $(QQ')$ inside the baryon is applied, but also the further fragmentation from the di-quark core to the baryon is involved\cite{doublyhadron}). For the proposal the observables evaluating the various asymmetries: the forward-backward one $A_{FB}$, the left-right one $A_{LR}$ and the combined forward-backward and left-right one $A^{FB}_{LR}$, are precisely defined. Additionally the way of how to calculate the asymmetry observables is depicted. In Section III, the theoretical uncertainties in computing the asymmetry observables $A_{FB}$, $A_{LR}$ and $A^{FB}_{LR}$ are shown, and it can be seen that the uncertainties for the observables are canceled or suppressed a lot. In order to see how sensitive to the effective mixing angles ${\rm sin^2\theta_{eff}^f}$ (f=lept,c,b) the asymmetry obsevables are, we also calculate and plot the results of the observables vs. various values of the effective mixing angles ${\rm sin^2\theta_{eff}^f}$ in proper manners. Section IV is reserved for discussing advantages and disadvantages of the proposal, and the conclusion on the proposal is made at the end.

\section{The production of the doubly heavy hadrons at a Z-factory and the asymmetries directly relating to the effective electro-weak mixing angles}
\label{Define}

In SM, the fermion $f$ couples to the gauge fields $A_{\mu}$ and $Z_{\mu}$ as follows:
\begin{eqnarray}
&V_{f \bar{f}\gamma}=e \sum_{f} Q_f \bar{\psi_f}\gamma^{\mu}\psi_f A_{\mu},\\
&V_{f \bar{f}Z}=\frac{e}{2{\rm cos}\theta_W{\rm sin}\theta_W} \sum_{f} Q_f \bar{\psi_f}\gamma^{\mu}(V^f-A^f\gamma^5)\psi_f Z_{\mu}
\end{eqnarray}
where $\psi_f$ means the field of the fermion with flavour $f$; $e(=g~{\rm sin}\theta_W)$ is the electric charge of a positron, and $Q_f$ is the charge of fermion $f$ in unit of e; $V^f$ and $A^f$,
\begin{eqnarray}
&V^f=t_{3L}(f)-2Q_f{\rm sin}^2\theta_W,\\
&A^f=t_{3L}(f),
\end{eqnarray}
are the operators for constructing the vector and axial-vector neutral currents for the fermion $f$ as in Eqs.(2,3,4),
and $t_{3L}(f)$ is the weak isospin of fermion $f$. Here the role of the electro-weak mixing angle played in the couplings can be seen clearly.

Considering higher order EW corrections, one may adopt the running QED coupling $\alpha(s)$ to compute the production, and replace the tree-level couplings of $Zf\bar{f}$ couplings with effective couplings $\bar{V}^f$ and $\bar{A}^f$\cite{EW}.
\begin{eqnarray}
&\bar{V}^f=t_{3L}(f)-2~Q_f~{\rm sin}^2\theta_{eff}^{f}\,, \nonumber\\
&\bar{A}^f=t_{3L}(f)\,.
\end{eqnarray}

\begin{figure}
\includegraphics[width=0.4\textwidth]{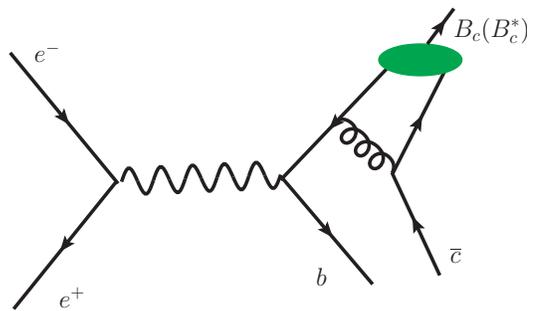}
\caption{One of the typical Feynman diagrams for the production of $B_c(B_c^*)$.} \label{feynman1}
\end{figure}

\begin{figure}
\includegraphics[width=0.4\textwidth]{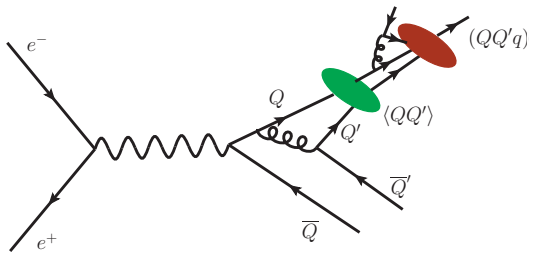}
\caption{One of the typical Feynman diagrams for the production of the doubly heavy baryon $(QQ'q)$.} \label{feynman2}
\end{figure}

One of the typical Feynman diagrams for the production of the $B_c(B_c^*)$ meson is shown in Fig.\ref{feynman1}, and one of the typical Feynman diagrams for the production of the $(QQ'q)$ baryon is shown in Fig.\ref{feynman2}.

The way to calculate the cross sections of the $B_c$ meson production, $e^+e^-\rightarrow (\gamma/Z) \rightarrow B_c (B^*_c)+b+\bar{c}$, can be found in Ref.\cite{doublyhadron,doublyhadron1}. Here we describe the calculation of the way of the doubly heavy baryon production in detail. In most references, such as \cite{cchhan,doublybaryon1,doublybaryon}, the doubly heavy baryon production is only about a relevant double heavy diquark core inside the doubly heavy baryon to be produced. In reference \cite{wuxg3}, the further process from the produced doubly heavy diquark core to the doubly heavy baryon is additionally considered. Here we calculate the doubly heavy baryon production as in reference\cite{wuxg3}.

Following Ref.\cite{wuxg3}, we treat the production of a doubly heavy baryon $(QQ'q)$ ($Q$, $Q'$ denote heavy
quarks, and $q$ denotes a light quark) by dividing it into two steps. The first step for the inclusive production of $(QQ'q)$: $e^+e^-\to (\gamma/Z) \to (QQ'q) +\bar{Q} +\bar{Q'}+\bar{q}$ is, similar to the $B_c$ inclusive production, the heavy diquark core $\langle QQ' \rangle_{\bar{3}}$ inside the baryon with suitable quantum numbers is produced as the process $e^+e^-\to (\gamma/Z) \to \langle QQ' \rangle_{\bar{3}} +\bar{Q} +\bar{Q'}$; the second step is, similar to the fragmentation of a heavy quark $\bar{Q}$ into a heavy meson ($\bar{Q}q$), the produced diquark core $\langle QQ' \rangle_{\bar{3}}$ fragments into the doubly heavy baryon ($QQ'q$) non-perturbatively by catching a light quark $q$ from the environment phenomenologically which is described by the so-called fragmentation function.

More precisely, now let us illustrate the way to compute the production of a doubly heavy baryon at a $Z$-factory.
Based on NRQCD, the cross section of the diquark core production can be written as:
\begin{eqnarray}
&d\sigma\left(e^+  e^-\rightarrow(\gamma/Z)\rightarrow\langle QQ' \rangle_{\bar{3}}+\bar{Q}+\bar{Q'} \right) \nonumber\\
&=\sum_n d\hat{\sigma} \left( e^+ e^-\to (\gamma/Z)\to (QQ')[n]+\bar{Q}+\bar{Q'}\right) \nonumber\\
&\cdot \langle \mathcal{O}^{\langle QQ'\rangle}(n)\rangle,
\end{eqnarray}
where the long-distance matrix element $\langle \mathcal{O}^{\langle QQ'\rangle}(n)\rangle$ represents the transition probability from a two quark state $(QQ')[n]$ into the diquark core state $\langle QQ'\rangle_{\bar{3}}$. Note that the wave function of the diquark core $\langle QQ'\rangle_{\bar{3}}$ should be totally antisymmetric under the exchange of the two quarks, so when $Q'=Q$, a diquark $\langle QQ\rangle_{\bar{3}}$ at the ground state (in $S$-wave and anti-color-triplet $\bar{\textbf{3}}$) must be in spin-triplet $S=1$.

As for the second step, the `inspired Peterson model'\cite{Peterson}\footnote{The original model is for a heavy quark to fragment to a heavy meson, whereas here we extend it for a doubly heavy diquark fragmenting into a baryon.} is employed on the fragmentation of the heavy diquark core $\langle QQ' \rangle_{\bar{3}}$ into the doubly heavy baryon $(QQ'q)$. According to the model, the fragmentation function is assumed as
\begin{equation}
D_{\langle QQ'\rangle_{\bar{3}}}^{H}(z)=\frac{N^H_{\langle QQ'\rangle_{\bar{3}}}}{z[1-1/z-\epsilon_{H}/(1-z)]^2},
\end{equation}
where the energy fraction $z\equiv\frac{E_{(QQ'q)}}{E_{\langle QQ'\rangle_{\bar{3}}}}$, $H$ is the doubly heavy baryon ($QQ'q$)
and $\epsilon_{H}\approx m_q^2/m_{\langle QQ'\rangle_{\bar{3}}}^2$. The normalization factor $N^H_{QQ'}$ is fixed by
\begin{equation}
\int_0^1 D_{\langle QQ' \rangle_{\bar{3}}}^{H}(z) dz=R_{\langle QQ' \rangle_{\bar{3}}}^{H},
\end{equation}
where $R_{\langle QQ' \rangle_{\bar{3}}}^{H}$ is fragmentation probability for the diquark core $\langle QQ'\rangle_{\bar{3}}$ to fragment into the doubly heavy baryon $H$. The fragmentation probabilities of the heavy diquark core to the relevant doubly heavy baryon with a specific light constituent quark may be estimated by the Lund Model\cite{pythia}, which gives $R_{\langle QQ' \rangle_{\bar{3}}}^{(QQ'u)}:R_{\langle QQ' \rangle_{\bar{3}}}^{(QQ'd)}:R_{\langle QQ' \rangle_{\bar{3}}}^{(QQ's)} \approx 1 : 1 : 0.3$. Furthermore, the contribution from the 'fragmentation', which catches  a heavy quark from the `environment' instead of a light quark, is very small, thus $R_{\langle QQ' \rangle_{\bar{3}}}^{(QQ'u)}=R_{\langle QQ' \rangle_{\bar{3}}}^{(QQ'd)}=43.5\%$ and $R_{\langle QQ' \rangle_{\bar{3}}}^{(QQ's)}=13.0\%$ are obtained. With these estimates on the probabilities, we can obtain the values of $N^H_{\langle QQ' \rangle_{\bar{3}}}$. In numerical calculations later on, the light quark masses $m_u=m_d=0.3 {\rm ~GeV}$ and $m_s=0.5 {\rm ~GeV}$ are adopted. Finally the production of the doubly heavy baryons may be calculated out via the convolution of the diquark differential cross section with the fragmentation function:
\begin{eqnarray}
&\displaystyle \frac{d\sigma(H(z))}{d z}=\int^1_z\frac{dy}{y}\frac{d\sigma(e^+e^-\to
\langle QQ' \rangle_{\bar{3}} (y)+\bar{Q}+\bar{Q'})}{d y}\nonumber\\
&\cdot D_{\langle QQ' \rangle_{\bar{3}}}^{H}(z/y)\;.
\end{eqnarray}

As shown in Ref.\cite{doublyhadron,doublyhadron1}, the differential cross-sections of the doubly heavy hadrons display asymmetrically. To suppress the uncertainties for the proposal as that in the tradition determinations, let us introduce $A_{FB}$, $A_{LR}$ and $A_{LR}^{FB}$, the observables of the various asymmetries, for the doubly heavy hadron production at a Z-factory. The observable $A_{FB}$ is for the forward-backward asymmetry and defined as
\begin{eqnarray}
A_{FB}=\frac{\sigma_F-\sigma_B}{\sigma_F+\sigma_B},
\end{eqnarray}
where $\sigma_F$ ($\sigma_B$) is the cross section for the produced doubly heavy-flavoured hadron travelling forward (backward) with respect to the direction of the initial electron.

If the collision beams are polarized, the observable $A_{LR}$ is for the left-right asymmetry and defined as
\begin{eqnarray}
A_{LR}=\frac{\sigma_L-\sigma_R}{\sigma_L+\sigma_R},
\end{eqnarray}
where $\sigma_L$ ($\sigma_R$) is the cross section for the initial left- (right-) handed electron. Moreover the observable $A_{LR}^{FB}$ is for the
combined left-right and forward-backward asymmetry and defined as
\begin{eqnarray}
A_{LR}^{FB}=\frac{\sigma_{LF}-\sigma_{LB}-\sigma_{RF}+\sigma_{RB}}{\sigma_{LF}+\sigma_{LB}+\sigma_{RF}+\sigma_{RB}}.
\end{eqnarray}
where $\sigma_{LF}$ and $\sigma_{RF}$ denote the production cross-sections in the forward direction with the initial electron left (right)-hand polarized, respectively, and $\sigma_{LB}$ and $\sigma_{RB}$ denote the production cross-sections in the backward direction with the initial electron left (right)-hand polarized respectively.

\section{Numerical values of the asymmetries and the sensitivities}
\label{numerical}

\begin{figure}
\includegraphics[width=0.4\textwidth]{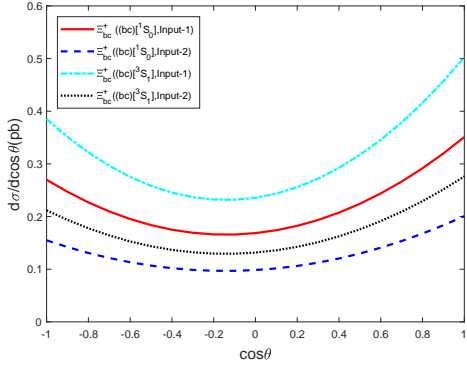}
\caption{Differential cross sections for the production of $\Xi_{bc}$ baryons with diquark cores in two possible spin multiplets, where``Input-1" and ``Input-2" are the two possible input masses of $b$-quark and $c$-quark.} \label{cosxibc}
\end{figure}

To calculate precise values of the asymmetry observables, the input parameters for numerical calculations were used as below:
\begin{eqnarray}
&& m_{_Z}=91.1876~{\rm GeV}\,,\nonumber\\
&&\alpha =1/129\,,\;\; \Gamma_{_Z}=2.4952~{\rm GeV}\,,
\end{eqnarray}
where $m_{_Z}$ is the mass of $Z$ boson, $\alpha=\alpha(m_{_Z})$ is the on-shell electromagnetic coupling constant, and $\Gamma_{_Z}$ is the width of $Z$ boson. The wave function at origin for the $B_c$ is adopted from \cite{pot} as follows:
\begin{eqnarray}
& &\vert R_S(0) \vert^2=1.642 ~{\rm GeV^3}.\nonumber \\
\end{eqnarray}

Since the production of the doubly heavy-flavored baryons is estimated by the two-step approach, the wave function at origin for the diquark cores, which is required in the approach, is taken from Ref.\cite{doublybaryon1} as follows:
\begin{eqnarray}
&&\vert \Psi_{\langle cc \rangle_{\bar{3}}}(0) \vert^2=0.039 ~{\rm GeV^3}\,,
\vert \Psi_{\langle bc \rangle_{\bar{3}}}(0) \vert^2=0.065 ~{\rm GeV^3}\,,\nonumber\\
&&\vert \Psi_{\langle bb \rangle_{\bar{3}}}(0) \vert^2=0.152 ~{\rm GeV^3}.
\end{eqnarray}
Here, the strong coupling constant is set as Ref.\cite{doublyhadron}, i.e., $\alpha_s(2 m_c)=0.237$ and $\alpha_s(2 m_b)=0.175$.
The mass values of the heavy quarks are tried with two possible choices as the inputs Eq.(\ref{masses}):
\begin{eqnarray}\label{masses}
&{\rm Input-1}:\;\;&m_b\simeq 4.9 ~{\rm GeV}, m_c\simeq 1.5 ~{\rm GeV}\,,\nonumber\\
&{\rm Input-2}:\;\;&m_b\simeq 5.1 ~{\rm GeV}, m_c\simeq 1.8 ~{\rm GeV}\,.
\end{eqnarray}

Below when a specific $f$-flavored effective electro-weak mixing angle ${\rm sin}^2\theta^f_{\rm eff}$ is highlighted for certain purposes of the proposal, then the way to set the value of the flavored effective electro-weak mixing angles is that the value of the specific $f$-flavored effective electro-weak mixing angle ${\rm sin}^2\theta^f_{\rm eff}$ is set as the requests of the purposes and the rest flavored effective electro-weak mixing angles ${\rm sin}^2\theta_{\rm eff}$, which appear in the calculation too, are set as the `common value' ${\rm sin}^2\theta^f_{\rm eff}=0.232$ as SM.

The differential cross sections for the production of the doubly heavy baryon $\Xi_{bc}$ were computed using the Input-1 or Input-2 masses given by Eq.\ref{masses}; the results are shown in Fig.\ref{cosxibc}. The difference between the differential cross sections with the two choices of the input heavy quark masses is quite large. Thus when any relevant fresh experimental data are available, then for the studies, including that for the proposal, the first work will be to suppress the uncertainty due to the input masses of the heavy quarks by fitting the fresh data so as to have input quark masses improved.

\begin{widetext}
\center
\begin{table}[h]
\begin{tabular}{ c c c c }
\hline\hline
~~Hadrons~~ & ~~~$A_{FB}$~~~ & ~~~~~~$A_{LR}$~~~~~~ & ~~~$A_{LR}^{FB}$~~~ \\
\hline
$B_c$(Input-1) & $(-9.07^{+5.03}_{-4.94})\times 10^{-2}$ & $0.143^{+0.078}_{-0.079}$ & $-0.634^{+0.000}_{-0.000}$ \\
$B_c$(Input-2) & $(-8.74^{+4.84}_{-4.76})\times 10^{-2}$ & $0.143^{+0.078}_{-0.079}$ & $-0.611^{+0.000}_{-0.000}$ \\
$B^*_c$ (Input-1) & $(-9.44^{+5.23}_{-5.14})\times 10^{-2}$ & $0.143^{+0.078}_{-0.079}$ & $-0.659^{+0.000}_{-0.000}$ \\
$B^*_c$ (Input-2)& $(-9.20^{+5.10}_{-5.01})\times 10^{-2}$ & $0.143^{+0.078}_{-0.079}$ & $-0.643^{+0.000}_{-0.000}$ \\
$(cc)[^3S_1]_{\bar{\textbf{3}}}$(Input-1) & $(7.03^{+3.83}_{-3.89})\times 10^{-2}$ & $0.143^{+0.078}_{-0.079}$ & $0.490^{+0.000}_{-0.000}$ \\
$(cc)[^3S_1]_{\bar{\textbf{3}}}$(Input-2) & $(7.00^{+3.81}_{-3.87})\times 10^{-2}$ & $0.143^{+0.078}_{-0.079}$ & $0.488^{+0.000}_{-0.000}$ \\
$(bc)[^1S_0]_{\bar{\textbf{3}}}$(Input-1)& $(9.40^{+5.11}_{-5.20})\times 10^{-2}$ & $0.143^{+0.078}_{-0.079}$ & $0.656^{+0.000}_{-0.000}$ \\
$(bc)[^1S_0]_{\bar{\textbf{3}}}$(Input-2)& $(9.21^{+5.03}_{-5.12})\times 10^{-2}$ & $0.143^{+0.078}_{-0.079}$ & $0.643^{+0.000}_{-0.000}$ \\
$(bc)[^3S_1]_{\bar{\textbf{3}}}$(Input-1) & $(9.64^{+5.25}_{-5.34})\times 10^{-2}$ & $0.143^{+0.078}_{-0.079}$ & $0.673^{+0.000}_{-0.000}$ \\
$(bc)[^3S_1]_{\bar{\textbf{3}}}$(Input-2) & $(9.52^{+5.18}_{-5.27})\times 10^{-2}$ & $0.143^{+0.078}_{-0.079}$ & $0.664^{+0.000}_{-0.000}$ \\
$(bb)[^3S_1]_{\bar{\textbf{3}}}$(Input-1) & $(8.97^{+4.88}_{-4.96})\times 10^{-2}$ & $0.143^{+0.078}_{-0.079}$ & $0.626^{+0.000}_{-0.000}$ \\
$(bb)[^3S_1]_{\bar{\textbf{3}}}$(Input-2) & $(8.91^{+4.86}_{-4.93})\times 10^{-2}$ & $0.143^{+0.078}_{-0.079}$ & $0.622^{+0.000}_{-0.000}$\\
\hline\hline
\end{tabular}
\caption{Calculated asymmetry observables when ${\rm sin}^2\theta_{\rm eff}^f=0.232$ (where f = b, c), the ranged electro-weak mixing angles are ${\rm sin}^2\theta_{\rm eff}^e=0.232\pm 0.010$, and the masses of the heavy quarks are taken as ``Input-1" and ``Input-2" in Eq.(\ref{masses}).}
\label{t01}
\end{table}
\begin{table}[h]
\begin{tabular}{ c c c c }
\hline\hline
~~Hadrons~~ & ~~~$A_{FB}$~~~ & ~~~~~~$A_{LR}$~~~~~~ & ~~~$A_{LR}^{FB}$~~~ \\
\hline
$B_c$(Input-1) & $(-9.07^{+0.05}_{-0.07})\times 10^{-2}$ & $0.143^{+0.000}_{-0.000}$ & $-0.634^{+0.005}_{-0.004}$ \\
$B_c$(Input-2) & $(-8.74^{+0.07}_{-0.07})\times 10^{-2}$ & $0.143^{+0.000}_{-0.000}$ & $-0.611^{+0.005}_{-0.005}$ \\
$B^*_c$(Input-1) & $(-9.44^{+0.08}_{-0.06})\times 10^{-2}$ & $0.143^{+0.000}_{-0.000}$ & $-0.659^{+0.005}_{-0.005}$ \\
$B^*_c$ (Input-2) & $(-9.20^{+0.07}_{-0.07})\times 10^{-2}$ & $0.143^{+0.000}_{-0.000}$ & $-0.643^{+0.005}_{-0.005}$ \\
$(bc)[^1S_0]_{\bar{\textbf{3}}}$(Input-1) & $(9.40^{+0.06}_{-0.07})\times 10^{-2}$ & $0.143^{+0.000}_{-0.000}$ & $0.656^{+0.004}_{-0.005}$ \\
$(bc)[^1S_0]_{\bar{\textbf{3}}}$(Input-2) & $(9.21^{+0.08}_{-0.06})\times 10^{-2}$ & $0.143^{+0.000}_{-0.000}$ & $0.643^{+0.05}_{-0.05}$ \\
$(bc)[^3S_1]_{\bar{\textbf{3}}}$(Input-1) & $(9.64^{+0.06}_{-0.07})\times 10^{-2}$ & $0.143^{+0.000}_{-0.000}$ & $0.673^{+0.004}_{-0.005}$ \\
$(bc)[^3S_1]_{\bar{\textbf{3}}}$(Input-2) & $(9.52^{+0.06}_{-0.07})\times 10^{-2}$ & $0.143^{+0.000}_{-0.000}$ & $0.664^{+0.04}_{-0.05}$ \\
$(bb)[^3S_1]_{\bar{\textbf{3}}}$(Input-1) & $(8.97^{+0.04}_{-0.05})\times 10^{-2}$ & $0.143^{+0.000}_{-0.000}$ & $0.626^{+0.003}_{-0.004}$ \\
$(bb)[^3S_1]_{\bar{\textbf{3}}}$(Input-2) & $(8.91^{+0.05}_{-0.04})\times 10^{-2}$ & $0.143^{+0.000}_{-0.000}$ & $0.622^{+0.03}_{-0.03}$ \\
\hline\hline
\end{tabular}
\caption{Calculated asymmetry observables when ${\rm sin}^2\theta_{\rm eff}^f=0.232$ (where f = lept,c), the ranged electro-weak mixing angles are ${\rm sin}^2\theta_{\rm eff}^b=0.232\pm 0.010$, and the masses of the heavy quarks are taken as ``Input-1" and ``Input-2" in Eq.(\ref{masses}).}
\label{t02}
\end{table}
\begin{table}[h]
\begin{tabular}{ c c c c }
\hline\hline
~~Hadrons~~ & ~~~$A_{FB}$~~~ & ~~~~~~$A_{LR}$~~~~~~ & ~~~$A_{LR}^{FB}$~~~ \\
\hline
$B_c$(Input-1) & $(-9.07^{+0.01}_{-0.02})\times 10^{-2}$ & $0.143^{+0.000}_{-0.000}$ & $-0.634^{+0.001}_{-0.001}$ \\
$B_c$(Input-2) & $(-8.74^{+0.02}_{-0.02})\times 10^{-2}$ & $0.143^{+0.000}_{-0.000}$ & $-0.611^{+0.001}_{-0.001}$ \\
$B^*_c$(Input-1) & $(-9.44^{+0.01}_{-0.01})\times 10^{-2}$ & $0.143^{+0.000}_{-0.000}$ & $-0.659^{+0.001}_{-0.001}$ \\
$B^*_c$ (Input-2) & $(-9.20^{+0.01}_{-0.01})\times 10^{-2}$ & $0.143^{+0.000}_{-0.000}$ & $-0.643^{+0.001}_{-0.001}$ \\
 $(cc)[^3S_1]_{\bar{\textbf{3}}}$(Input-1) & $(7.03^{+0.35}_{-0.37})\times 10^{-2}$ & $0.143^{+0.000}_{-0.000}$ & $0.490^{+0.025}_{-0.026}$ \\
 $(cc)[^3S_1]_{\bar{\textbf{3}}}$(Input-2) & $(7.00^{+0.35}_{-0.37})\times 10^{-2}$ & $0.143^{+0.000}_{-0.000}$ & $0.488^{+0.025}_{-0.026}$ \\
$(bc)[^1S_0]_{\bar{\textbf{3}}}$(Input-1) & $(9.40^{+0.01}_{-0.01})\times 10^{-2}$ & $0.143^{+0.000}_{-0.000}$ & $0.656^{+0.001}_{-0.001}$ \\
$(bc)[^1S_0]_{\bar{\textbf{3}}}$(Input-2) & $(9.21^{+0.02}_{-0.01})\times 10^{-2}$ & $0.143^{+0.000}_{-0.000}$ & $0.643^{+0.001}_{-0.001}$ \\
$(bc)[^3S_1]_{\bar{\textbf{3}}}$(Input-1) & $(9.641^{+0.004}_{-0.005})\times 10^{-2}$ & $0.143^{+0.000}_{-0.000}$ & $0.6727^{+0.0003}_{-0.0003}$ \\
$(bc)[^3S_1]_{\bar{\textbf{3}}}$(Input-2) & $(9.520^{+0.007}_{-0.008})\times 10^{-2}$ & $0.143^{+0.000}_{-0.000}$ & $0.6641^{+0.0005}_{-0.0006}$ \\
\hline\hline
\end{tabular}
\caption{Calculated asymmetry observables when ${\rm sin}^2\theta_{\rm eff}^f=0.232$ (here f = lept,b), the ranged electro-weak mixing angles are ${\rm sin}^2\theta_{\rm eff}^c=0.232\pm 0.010$, and the masses of the heavy quarks are taken as ``Input-1" and ``Input-2" in Eq.(\ref{masses}).}
\label{t03}
\end{table}

\end{widetext}

In the production formulas, the strong coupling $\alpha_s$ and the wave function at origin both appear as independent factors, so according to the definition of the asymmetry observables, they are cancelled in the observables, i.e., the input values of $\alpha_s$ and the wave function at origin do not affect the observables at all. Thus the input quark masses contribute to the theoretical uncertainty primarily when determining the effective electro-weak mixing angles by measuring the asymmetry observables.

To see the impact of the input masses of the quarks and the effective mixing angles on the asymmetry observables $A_{FB}$, $A_{LR}$ and $A^{FB}_{LR}$ of the produced hadron, the simulation was performed using two possible choice of the quark mass inputs and with the flavored effective mixing angles taken as the common value $0.232$ exactly, except one of the angles with a variation ($\pm 0.010$); three cases were thus trialed: i) ${\rm sin}^2\theta_{\rm eff}^e=0.232\pm 0.010$, ${\rm sin}^2\theta_{\rm eff}^{\rm b,c}=0.232$, ii) ${\rm sin}^2\theta_{\rm eff}^{\rm b}=0.232\pm 0.010$, ${\rm sin}^2\theta_{\rm eff}^{\rm e,c}=0.232$ and iii). ${\rm sin}^2\theta_{\rm eff}^{\rm c}=0.232\pm 0.010$, ${\rm sin}^2\theta_{\rm eff}^{\rm e,b}=0.232$. The obtained results of cases i -- iii are shown in tables \ref{t01},\ref{t02},\ref{t03}, respectively. The obtained variations of the resultant asymmetry observables are due to the input `variation' of the electro-weak mixing angles in the cases i), ii), and iii) respectively. Using Input-1 or Input-2 as the masses resulted in similar asymmetry observables, indicating that the theoretical uncertainty caused by the input masses of the heavy quarks was also suppressed substantially. However, when more and precise experimental data on spectra, production and decays of the doubly heavy hadrons are available, the uncertainty range of the input mass parameters (e.g., that from ``Input-1" to ``Input-2") must be narrowed by fitting all the fresh available data, so with the narrowed uncertainty range of the input mass parameters, the full uncertainties on the effective mixing angles would accordingly decrease.

\begin{widetext}

\begin{table}
\begin{tabular}{ c c c}
\hline\hline
~~Hadrons~~ & ~~$\sigma$(pb)~~& Events /Year \\
\hline
$B_c$ & 2.69 & $2.69\times 10^7$ \\
$B^*_c$ & 3.76 & $3.76 \times 10^7$ \\
$\Xi_{cc}^+(\Xi_{cc}^{++})$ & 0.395 & $3.95\times 10^6$\\
$\Omega_{cc}^+$ & 0.118 & $1.18 \times 10^6$\\
$\Xi_{bc}^0(\Xi_{bc}^+)((bc)[^1S_0]_{\bar{\textbf{3}}})$ & 0.432 & $4.32\times 10^6$ \\
$\Omega_{bc}((bc)[^1S_0]_{\bar{\textbf{3}}})$ & 0.129 & $1.29\times 10^6$\\
$\Xi_{bc}^0(\Xi_{bc}^+)((bc)[^3S_1]_{\bar{\textbf{3}}})$ & 0.610 & $6.10 \times 10^6$\\
$\Omega_{bc}((bc)[^3S_1]_{\bar{\textbf{3}}})$ & 0.183 & $1.83\times 10^6$\\
$\Xi_{bb}^-(\Xi_{bb}^0)$ & $2.34\times 10^{-2}$ & $2.34\times 10^5$\\
$\Omega_{bb}^-$ & $0.70 \times 10^{-2}$ & $7.00\times 10^4$\\
\hline\hline
\end{tabular}
\caption{The total cross sections $\sigma$ are those of the doubly heavy hadron production when ${\rm sin}^2\theta_{\rm eff}^f=0.232$ (f = lept,c,b). The `Number of Events' are event numbers a year of the produced hadrons at such a Z-factory with luminosity $10^{36} cm^{-2}s^{-1}$ and operating longer than $10^7s$. The estimations for the production were made assuming input quark masses of $m_b=4.9\,{\rm GeV}$, $m_c=1.5\,{\rm GeV}$, $m_u\simeq m_d\simeq 0.3{\rm GeV}$, and $m_s\simeq 0.5{\rm GeV}$.}
\label{tasy000}
\end{table}

\end{widetext}

The cross sections were calculated and the numbers of events of the doubly heavy-flavored hadrons produced at a super $Z$-factory were estimated; the results are summarized in Table. \ref{tasy000}. Here, the Input-1 case was used as the input masses of the heavy quarks, and the masses of the light quarks in the doubly heavy baryons were taken as $m_u\simeq m_d\simeq 0.3{\rm GeV}, m_s\simeq 0.5{\rm GeV}$. The luminosity of the considered $Z$-factory was assumed as $10^{36} cm^{-2}s^{-1}$ and the data-taking time at the factory in a year was assumed as $10^7 s$. Approximately $10^4 \sim 10^7$ events/year for the various doubly heavy-flavored hadrons listed in Table. \ref{tasy000} can be collected at such a super $Z$-factory.


To see how sensitive the asymmetry observables $A_{FB}$, $A_{LR}$ and $A_{LR}^{FB}$ of the doubly-heavy hadron production are to the values of the effective specific-flavored mixing angles ${\rm sin}^2\theta_{\rm eff}^f$ (f=lept;\, c, b), the asymmetry observables of the production were calculated under the cases where only one of the effective flavored mixing angles varies but the rest relevant mixing angles are fixed by the value 0.232, and the results are presented in Figs.\ref{afbe},\ref{alre},\ref{alrfbe} accordingly.

\begin{widetext}

\begin{figure}[!h]
\centering
\includegraphics[width=0.3\textwidth]{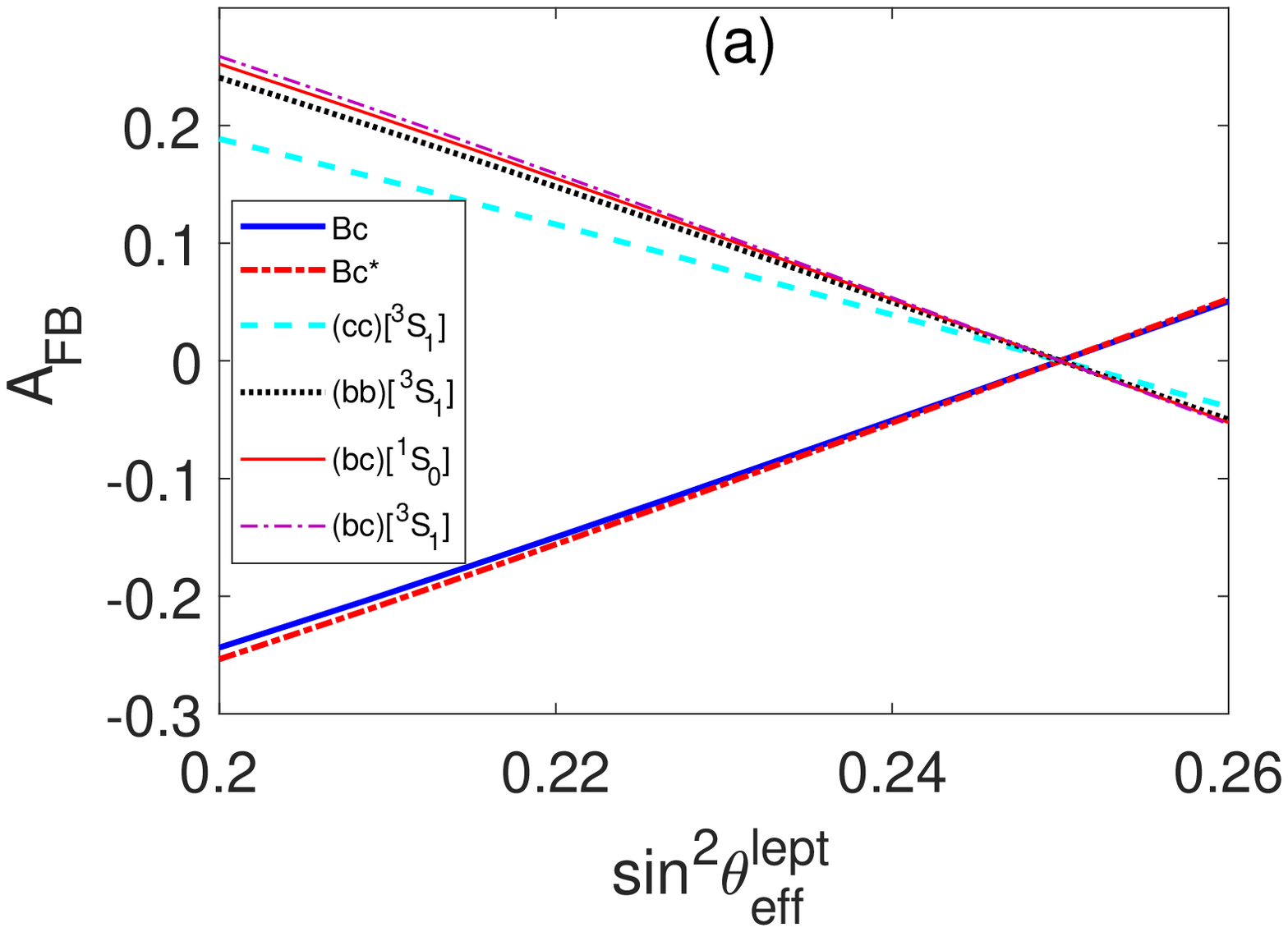}
\includegraphics[width=0.3\textwidth]{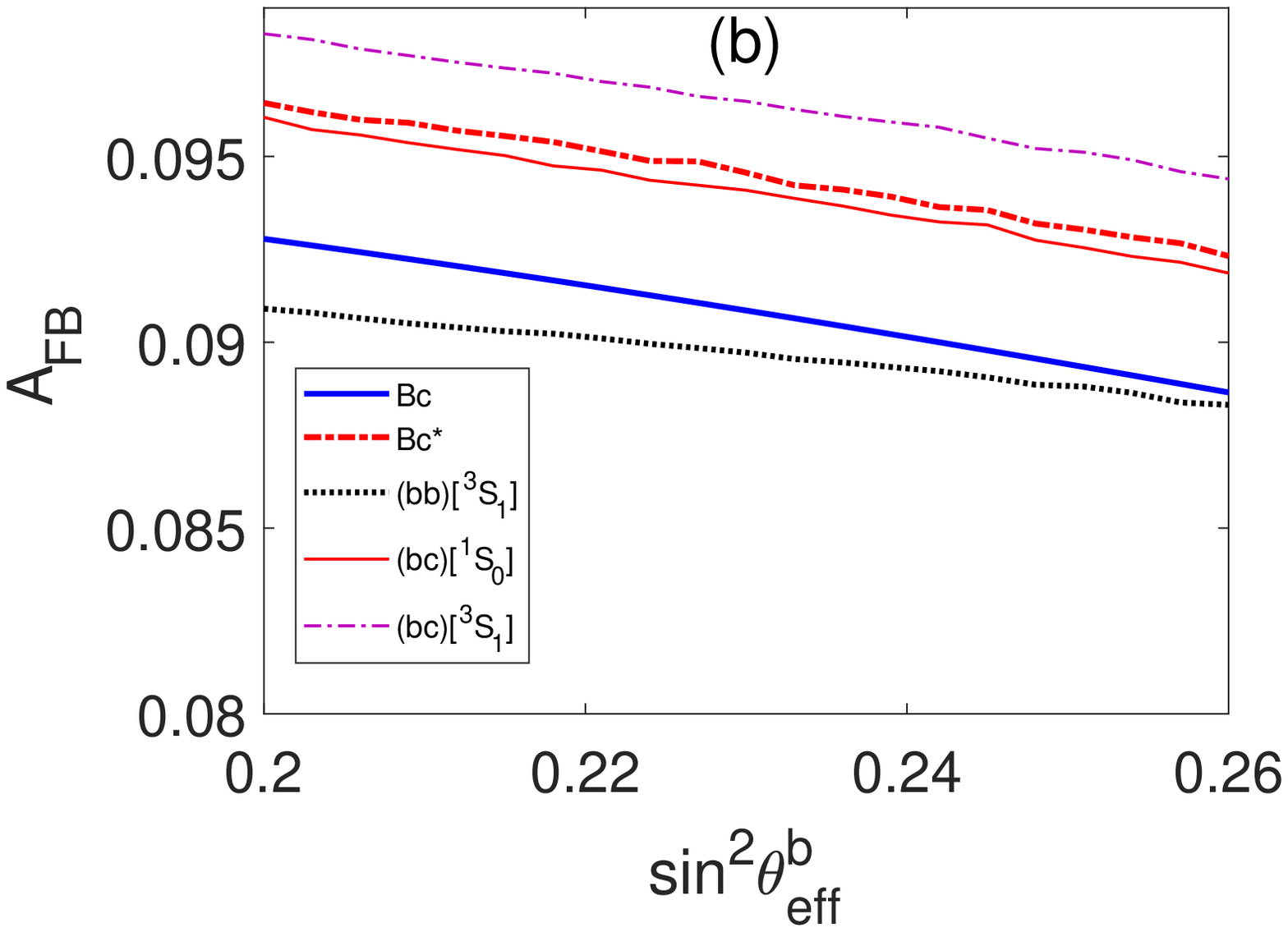}
\includegraphics[width=0.3\textwidth]{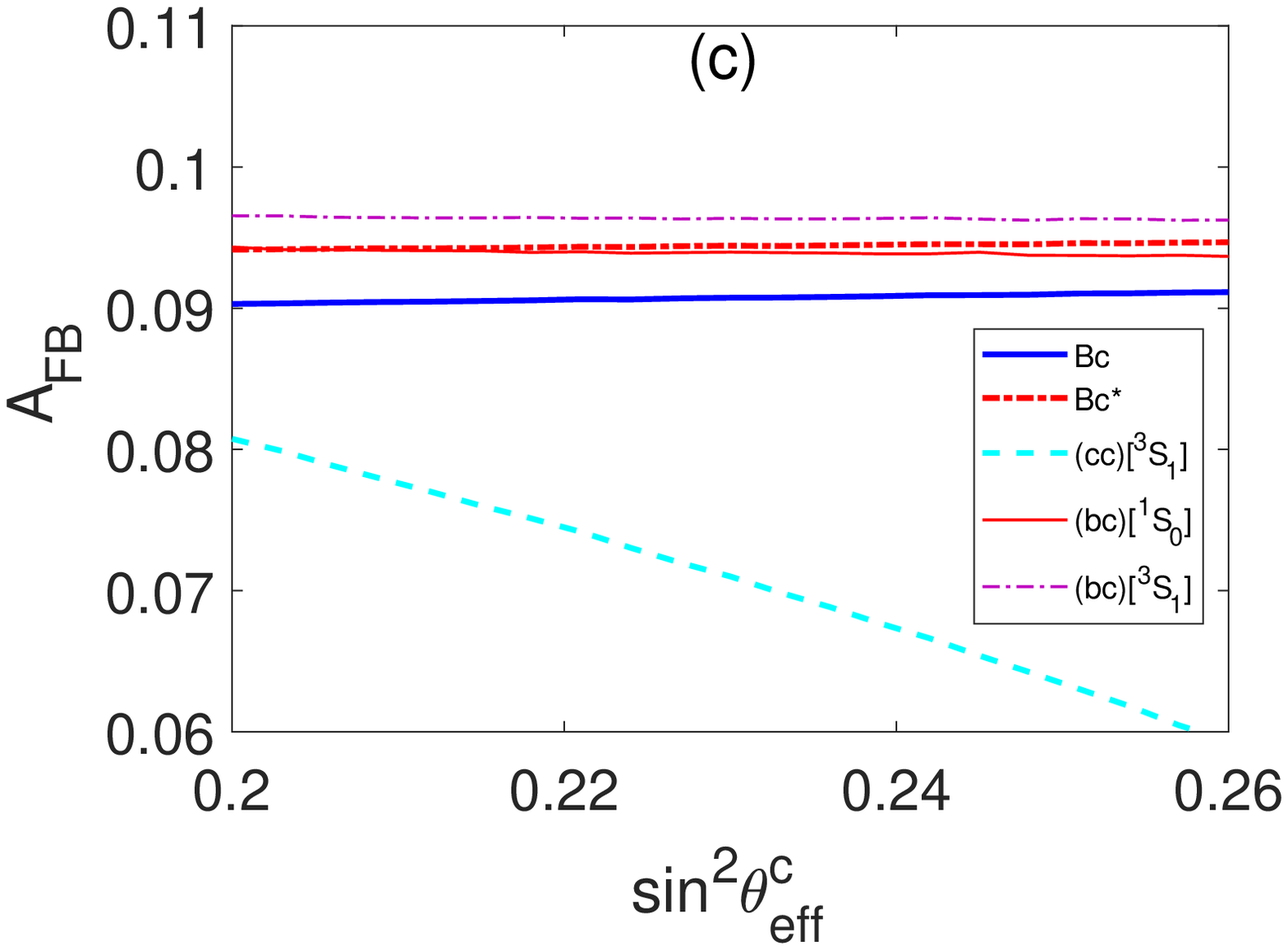}
\caption{ The forward--backward asymmetries of the production of the doubly heavy-flavored hadrons as functions of {\bf (a)} ${\rm sin}^2\theta^{\rm lept}_{\rm eff}$, with the relevant ${\rm sin}^2\theta^{b}_{\rm eff}$ and ${\rm sin}^2\theta^{c}_{\rm eff}$ fixed at 0.232; {\bf (b)} ${\rm sin}^2\theta^{b}_{\rm eff}$, with the relevant ${\rm sin}^2\theta^{\rm lept}_{\rm eff}$ and ${\rm sin}^2\theta^{\rm c}_{\rm eff}$ fixed at 0.232; and {\bf (c)} ${\rm sin}^2\theta^{c}_{\rm eff}$, with the relevant ${\rm sin}^2\theta^{\rm lept}_{\rm eff}$ and ${\rm sin}^2\theta^{\rm b}_{\rm eff}$ fixed at 0.232. To fit the results into one figure, the $B_c$ and $B^*_c$ were multiplied by a factor of 1 in (b) and (c).} \label{afbe}
\end{figure}

\begin{figure}[!h]
\centering
\includegraphics[width=0.3\textwidth]{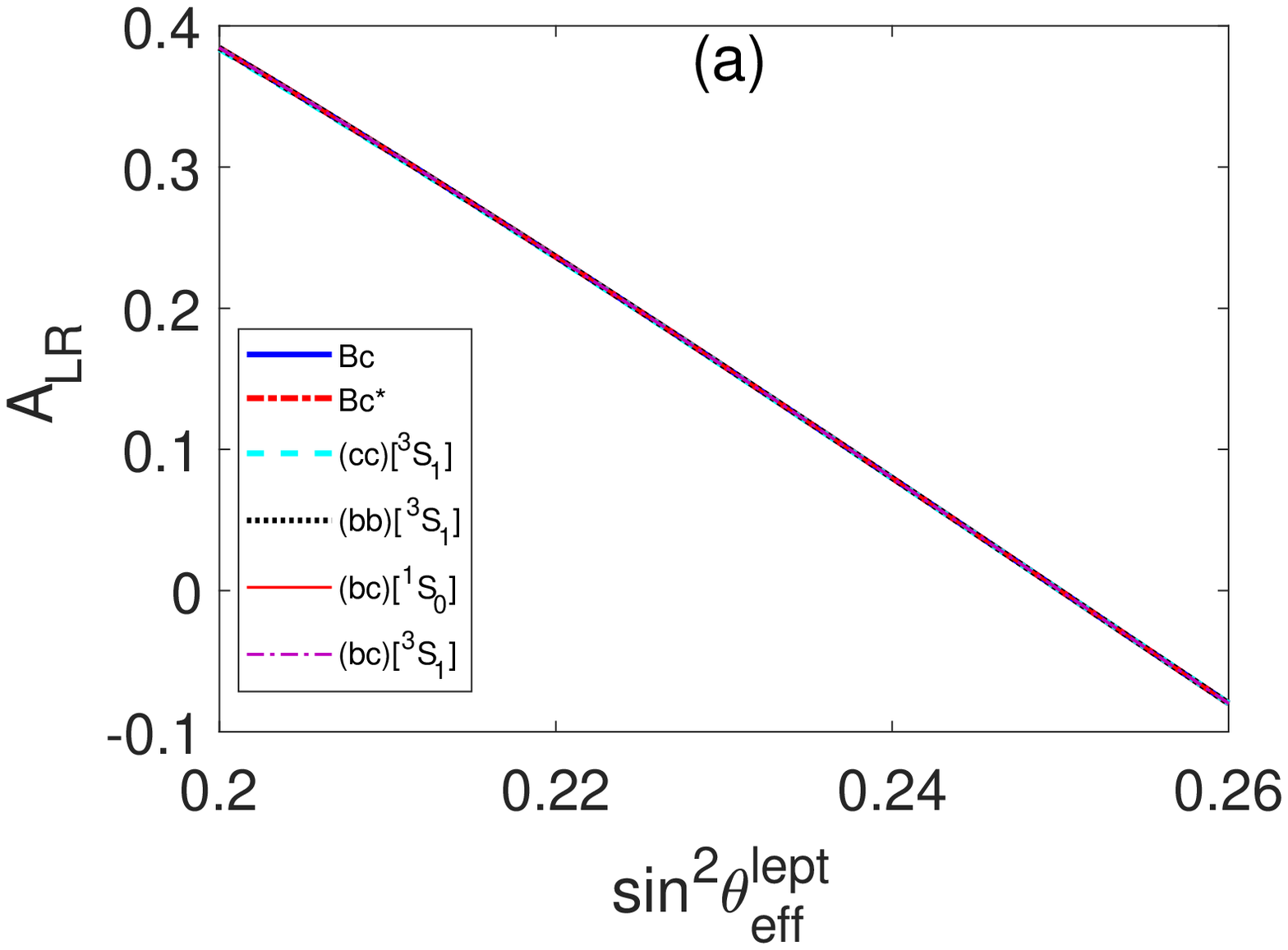}
\includegraphics[width=0.3\textwidth]{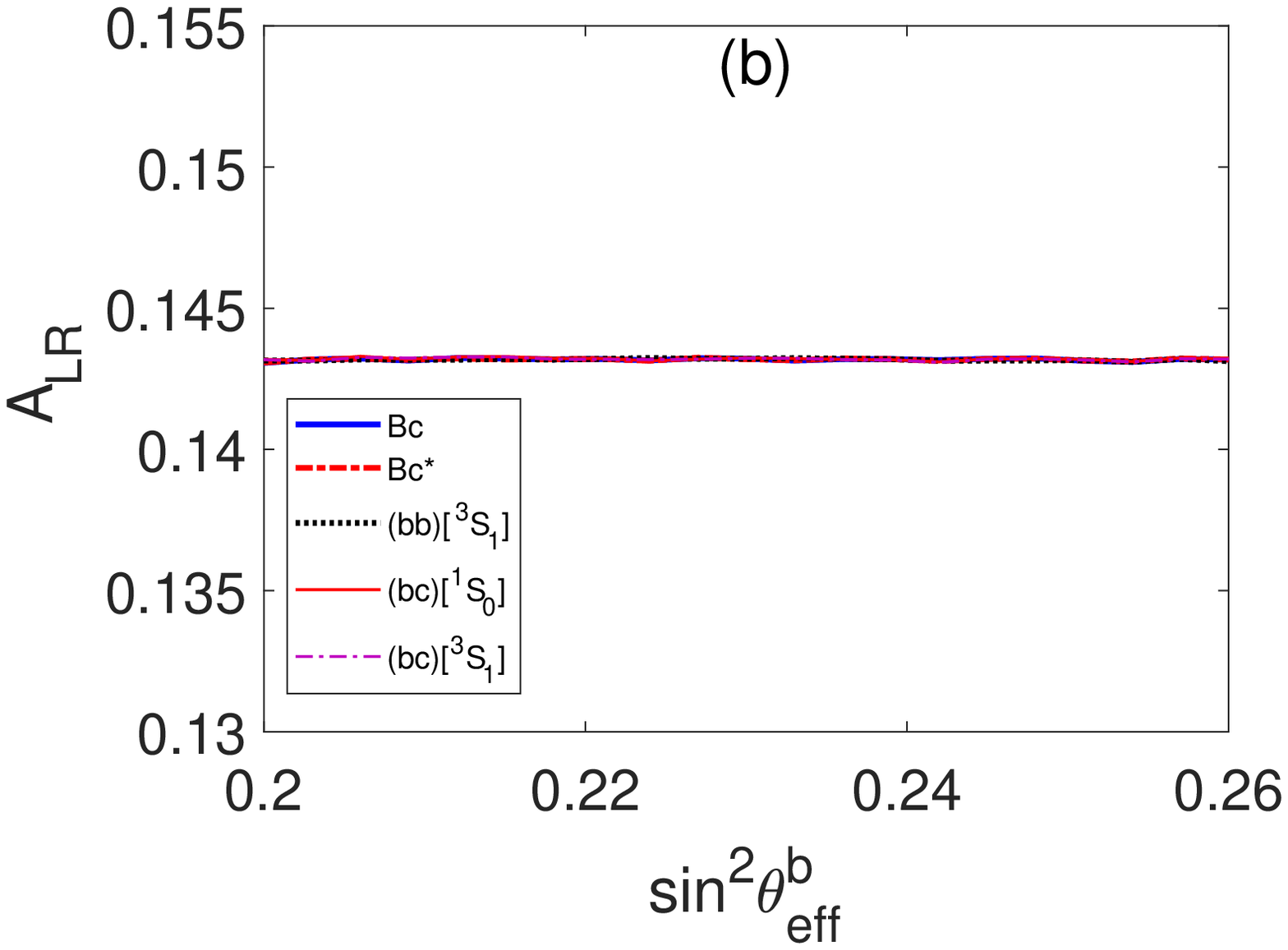}
\includegraphics[width=0.3\textwidth]{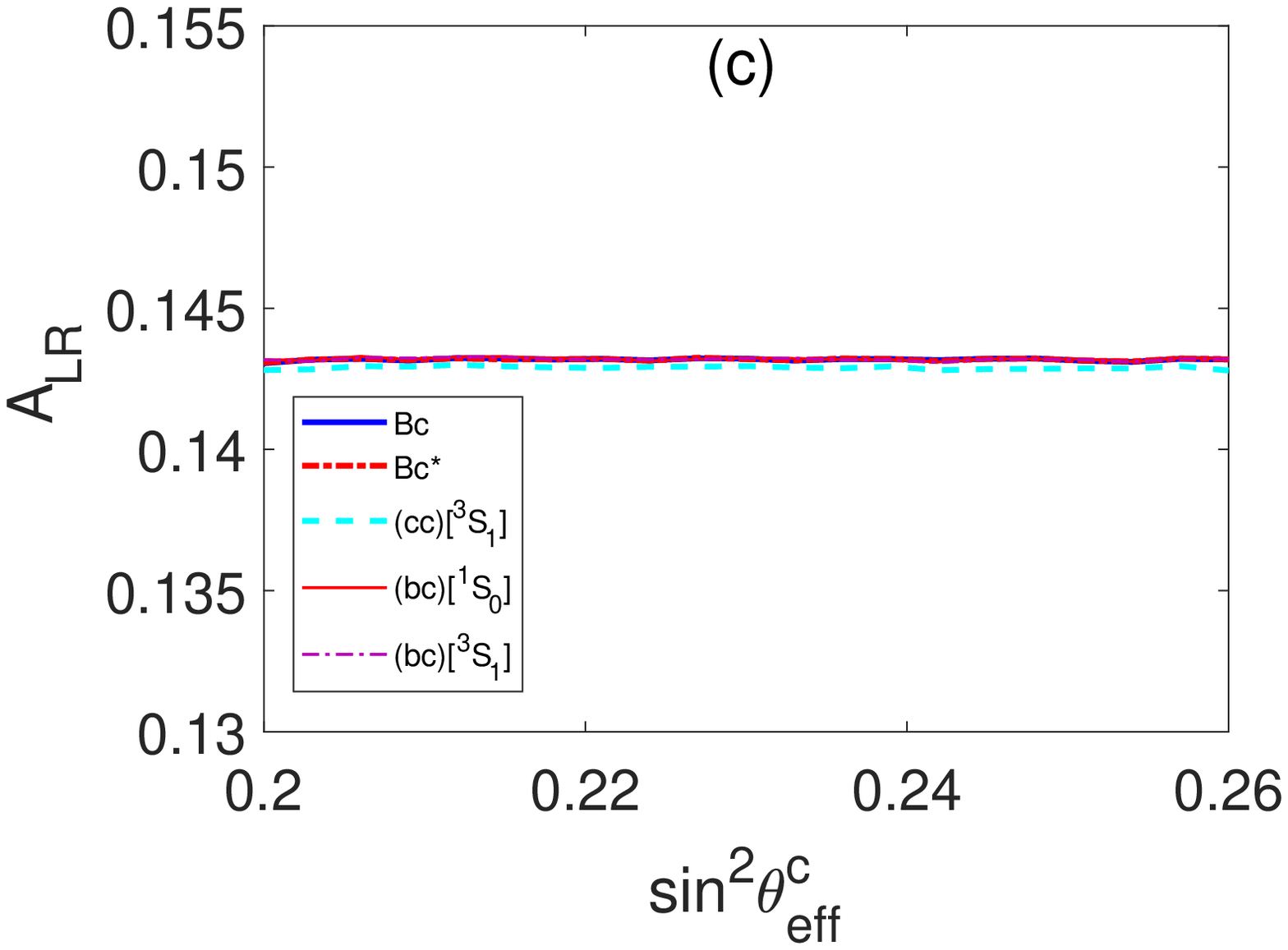}
\caption{The left--right asymmetries of the production of the doubly heavy-flavored hadrons as functions of {\bf (a)} ${\rm sin}^2\theta^{\rm lept}_{\rm eff}$ (lept=e), with the relevant ${\rm sin}^2\theta^{b}_{\rm eff}$ and ${\rm sin}^2\theta^{c}_{\rm eff}$ fixed at 0.232; {\bf (b)} ${\rm sin}^2\theta^{b}_{\rm eff}$, with the relevant ${\rm sin}^2\theta^{\rm lept}_{\rm eff}$ and ${\rm sin}^2\theta^{\rm c}_{\rm eff}$ fixed at 0.232; and {\bf (c)} ${\rm sin}^2\theta^{c}_{\rm eff}$, with the relevant ${\rm sin}^2\theta^{\rm lept}_{\rm eff}$ and ${\rm sin}^2\theta^{\rm b}_{\rm eff}$ fixed at 0.232. To fit the results into one figure, the $B_c$ and $B^*_c$ were multiplied by a factor of ?1 in (b) and (c).} \label{alre}
\end{figure}

\begin{figure}[!h]
\centering
\includegraphics[width=0.3\textwidth]{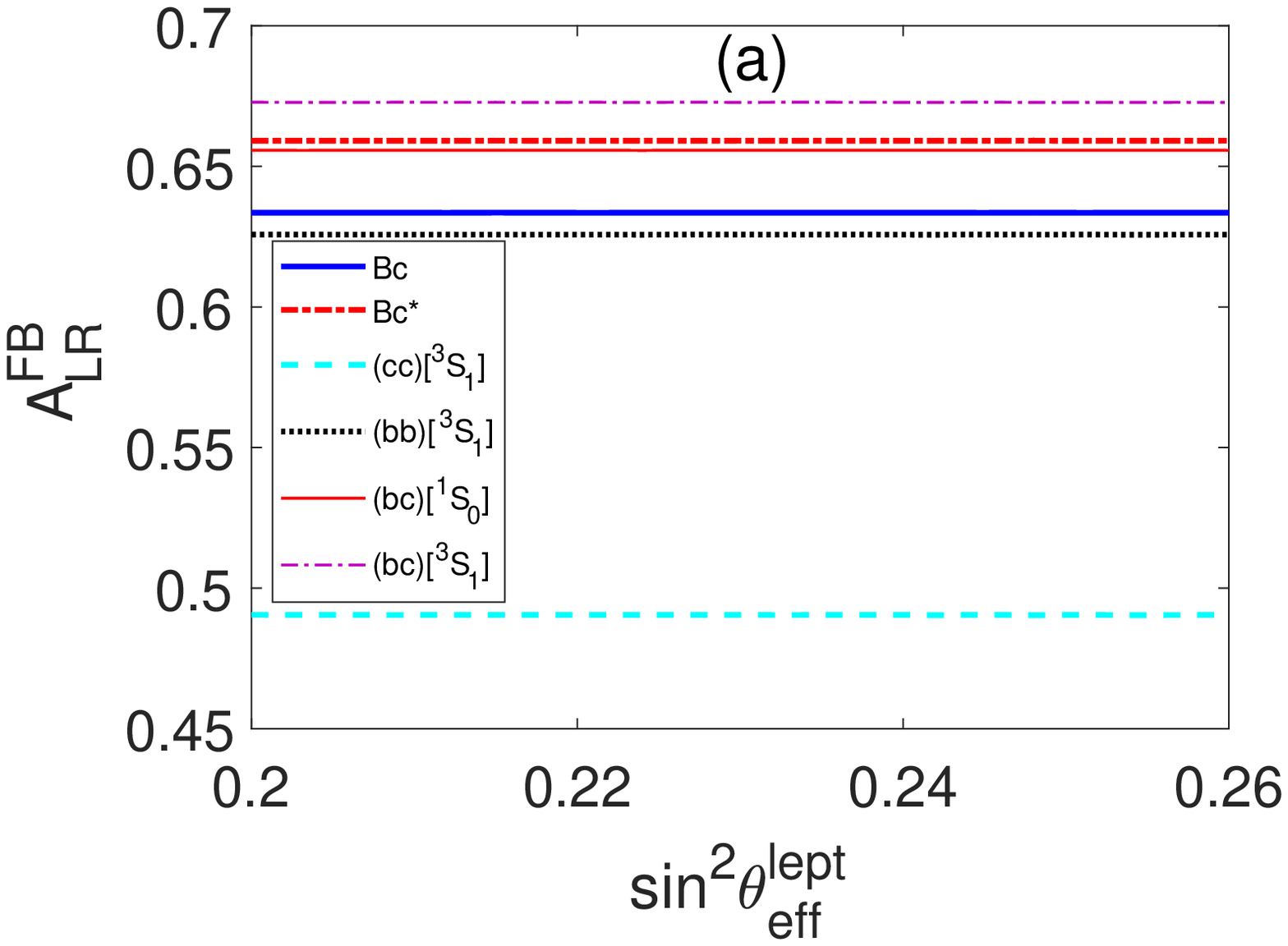}
\includegraphics[width=0.3\textwidth]{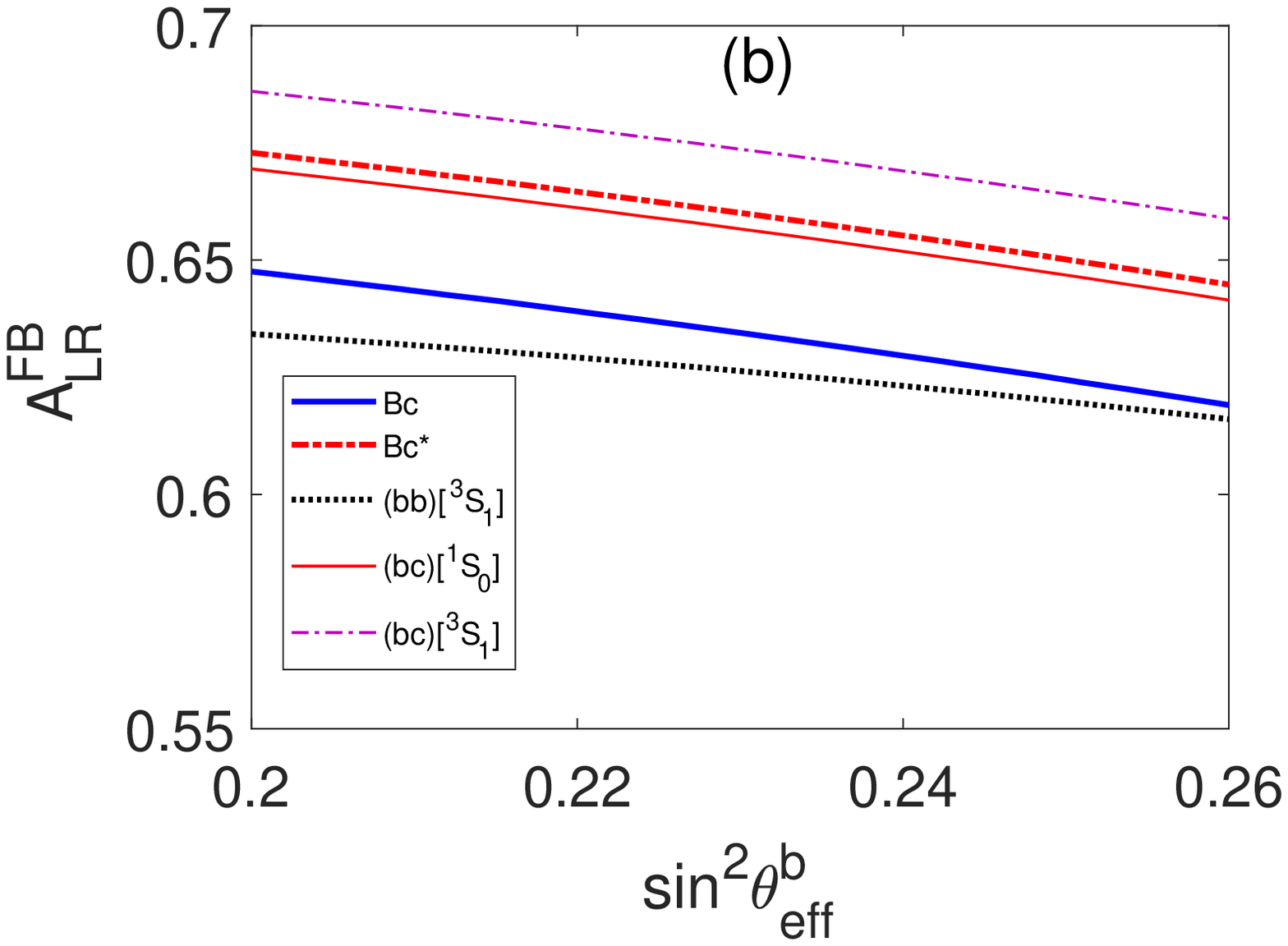}
\includegraphics[width=0.3\textwidth]{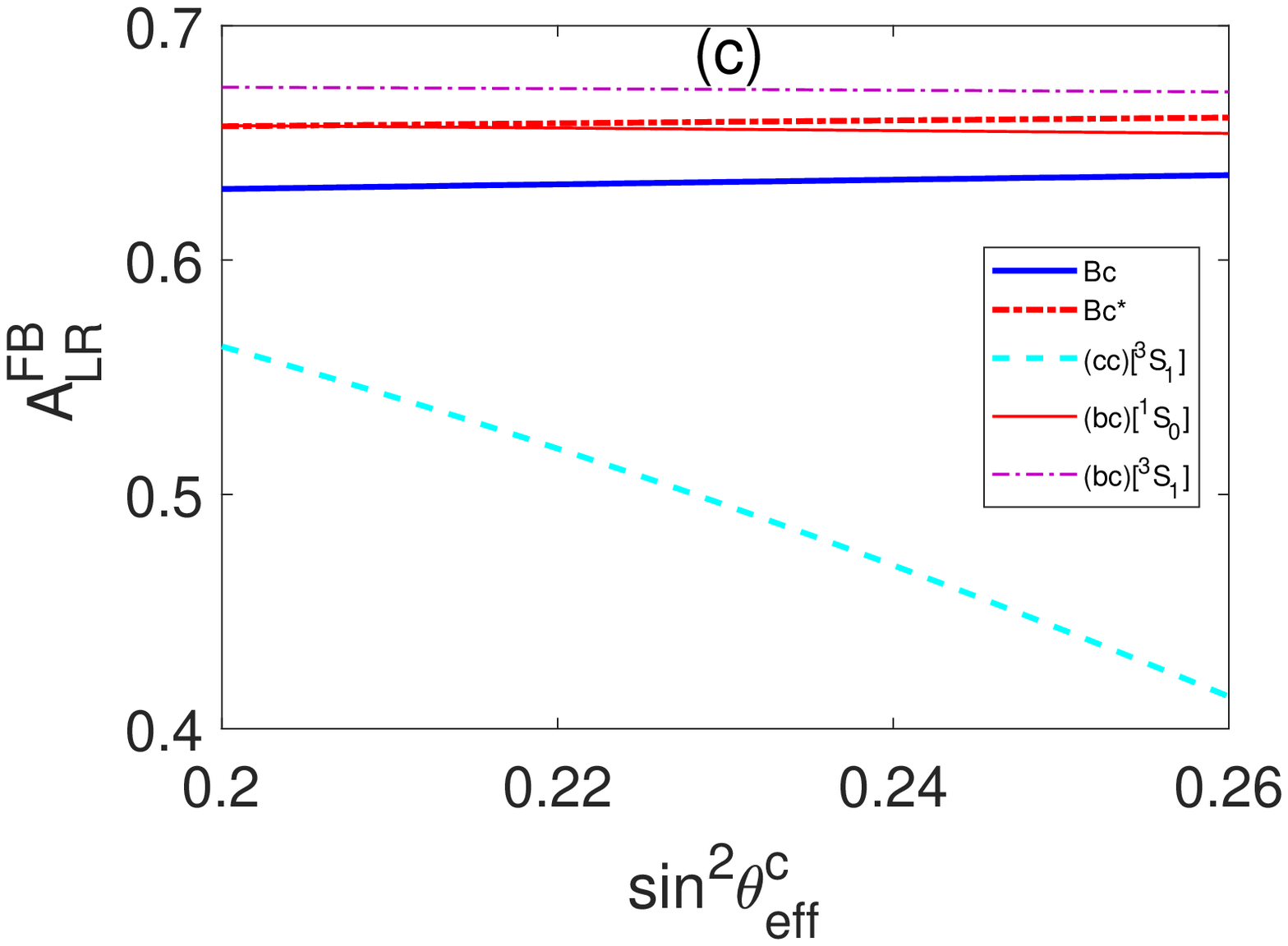}
\caption{The left--right forward--backward asymmetries of the production of the doubly heavy-flavored hadrons as functions of {\bf (a)} ${\rm sin}^2\theta^{\rm lept}_{\rm eff}$, with the relevant ${\rm sin}^2\theta^{b}_{\rm eff}$ and ${\rm sin}^2\theta^{c}_{\rm eff}$ fixed at 0.232; {\bf (b)} ${\rm sin}^2\theta^{b}_{\rm eff}$, with the relevant ${\rm sin}^2\theta^{\rm lept}_{\rm eff}$ and ${\rm sin}^2\theta^{\rm c}_{\rm eff}$ fixed at 0.232; and {\bf (c)} ${\rm sin}^2\theta^{c}_{\rm eff}$, with the relevant ${\rm sin}^2\theta^{\rm lept}_{\rm eff}$ and ${\rm sin}^2\theta^{\rm b}_{\rm eff}$ fixed at 0.232. To fit the results into one figure, the $B_c$ and $B^*_c$ were multiplied by a factor of 1 in (b) and (c).} \label{alrfbe}
\end{figure}

\begin{figure}[!h]
\centering
\includegraphics[width=0.3\textwidth]{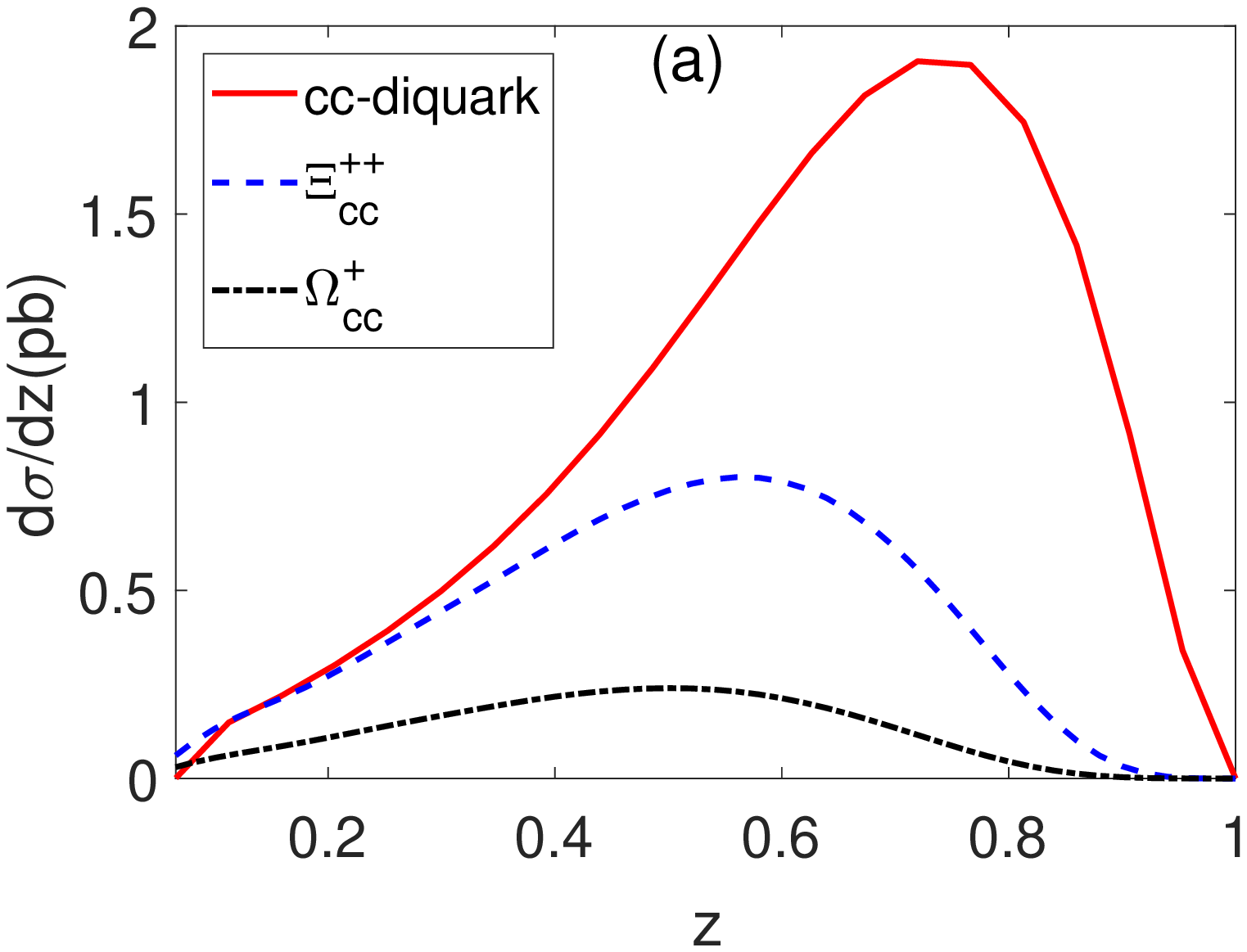}
\includegraphics[width=0.3\textwidth]{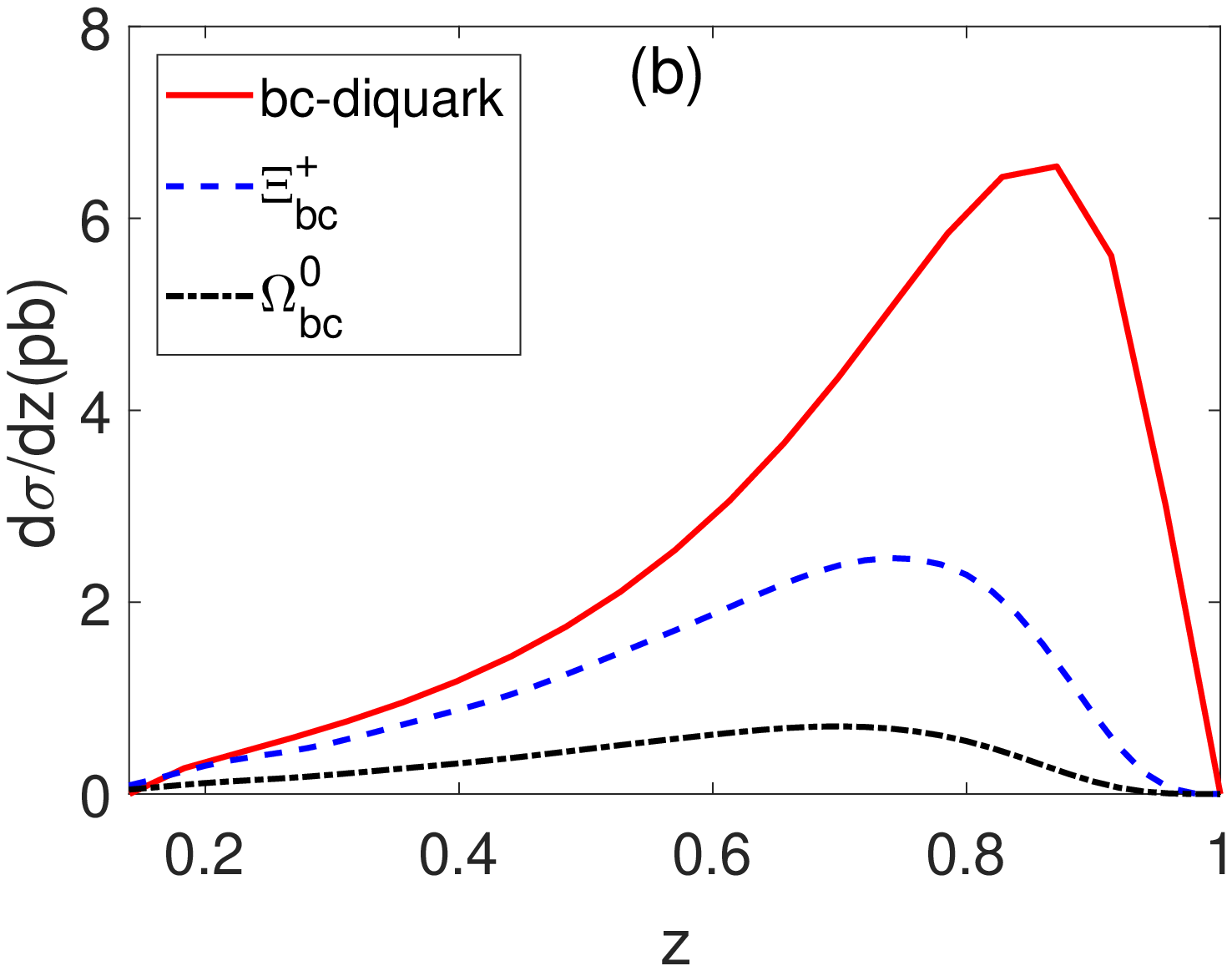}
\includegraphics[width=0.3\textwidth]{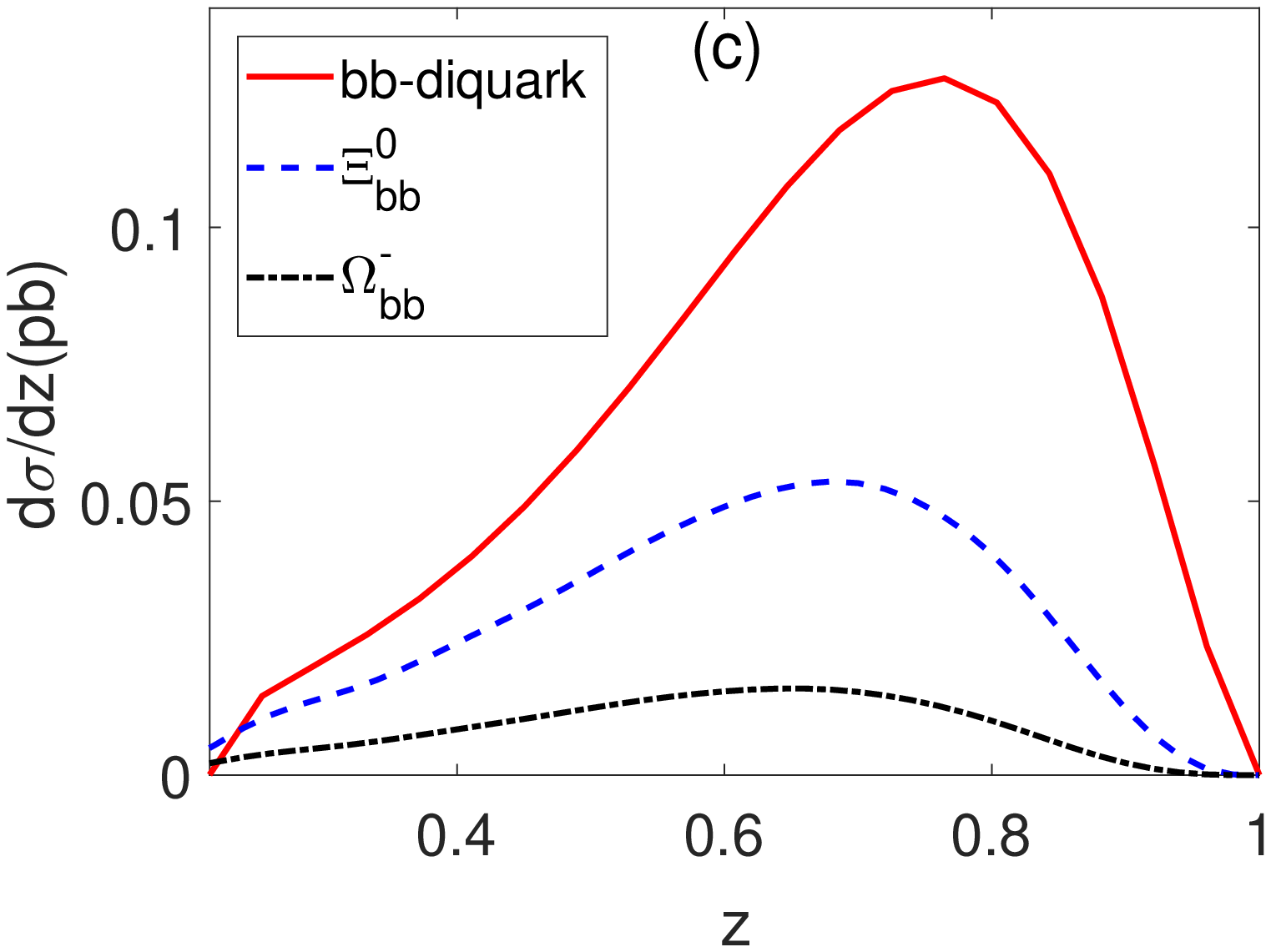}
\caption{The differential energy distribution for {\bf (a)} doubly charmed baryons $(ccq)$, {\bf (b)} bottom-charmed baryons $(bcq)$, and {\bf (c)} doubly bottomed baryons $(bbq)$ with different light constituent quarks.} \label{zcc}
\end{figure}

\end{widetext}

Since a doubly heavy baryon contains a light quark as well as a heavy diquark core, the differential energy distributions for the produced doubly heavy baryons at a $Z$-factory are also interesting as references for the proposal. Thus, the energy distributions were calculated, and the resulting differential energy distributions $d\sigma/dz$ for the doubly heavy baryons and for the $\langle QQ'\rangle$-diquark core are presented in Fig.\ref{zcc}. The distributions of the doubly heavy baryons are obviously changed in comparison with the distributions of the corresponding diquark core. The maximum point of the distributions of the doubly heavy baryons $(QQ'q)$ is shifted to a smaller value of the energy fraction in comparison with that of the distributions of the corresponding diquark core $\langle QQ'\rangle$.





\section{Discussion and Conclusions}
\label{conclusion}

To test SM precisely and to find clues deviating from SM, the production of the doubly heavy-flavored hadrons ($B_c$ meson, $\Xi_{cc},\Xi_{bc}, \Xi_{bb}\cdots$ baryons) at a super $Z$ factory (an $e^+e^-$ collider running around $Z$-resonance) was studied. The capability and possibility in determining the effective electro-weak mixing angles ${\rm sin}^2\theta^{\rm lept}_{\rm eff}$ (lept=e), ${\rm sin}^2\theta_{\rm eff}^b$ and ${\rm sin}^2\theta_{\rm eff}^c$ by measuring the asymmetry observables $A_{FB}$, $A_{LR}$ and $A^{FB}_{LR}$ of the relevant produced doubly heavy-flavored hadrons at such a super $Z$-factory were highlighted. The results of the studies are presented in tables.\ref{t01},\ref{t02},\ref{t03} and Figs.\ref{afbe},\ref{alre},\ref{alrfbe}. The results clearly show that the suggested asymmetry observables $A_{FB}$, $A_{LR}$ and $A^{FB}_{LR}$ of the produced doubly heavy hadrons are sensitive to the effective mixing angles, especially, of the flavors $lept, c, b$, and the theoretical uncertainties of the observables, such as those caused by the `wave functions at origin' and the strong coupling $\alpha_S$, are cancelled, even those caused by the input heavy quark masses are suppressed greatly. For the traditional determinations on the effective electro-weak mixing angles at LEP-I and SLC, due to the possible jet-flavor misidentification, the obscured jet-direction of the produced quark by fragmentation of the quark into the jet, etc. the experimental systematic errors on the effective electro-weak mixing angles are `hard', i.e., they cannot be easy to be suppressed further, additionally, there were some inconsistences among the determined effective mixing angles ${\rm sin}^2\theta^{\rm f}_{\rm eff}$ (f=c,b; e) obtained by the tradition determinations. Thus here, it was proposed that, similar to the tranditional method, to determine the effective mixing angles ${\rm sin}^2\theta^{\rm f}_{\rm eff}$ (f=c,b; e) by measuring the asymmetry observables $A_{FB}$, $A_{LR}$, and $A^{FB}_{LR}$ of the inclusively produced doubly heavy hadrons. From Figs.\ref{afbe},\ref{alre},\ref{alrfbe}, it can be seen that when the beams are polarized, the forward-backward, left--right and left--right forward--backward asymmetry observables of the $\Xi_{cc}, \Omega_{cc}$ ($\Xi_{bb}, \Omega_{bb}$) production at a super $Z$-factory are more ideal in the determination of the flavored mixing angles ${\rm sin}^2\theta^{c}_{\rm eff}$ (${\rm sin}^2\theta^{b}_{\rm eff}$) than the asymmetry observables of the production of other doubly heavy hadrons. This is likely because the production of the forward--backward asymmetry observable $A_{FB}$ and the left--right forward--backward asymmetry observables $A^{FB}_{LR}$ of $\Xi_{cc}, \Omega_{cc}$ ($\Xi_{bb}, \Omega_{bb}$) precisely depend on ${\rm sin}^2\theta^{\rm e}_{\rm eff}$ and ${\rm sin}^2\theta^{\rm c}_{\rm eff}$ (${\rm sin}^2\theta^{\rm b}_{\rm eff}$), whereas the left--right asymmetry observable $A_{LR}$ directly depends on ${\rm sin}^2\theta^{\rm lept}_{\rm eff}$.

The advantages of the proposal in determining the effective mixing angles are that the produced doubly heavy flavor(s) and the out-going direction of the produced doubly-heavy hadrons, which are crucial for the determination, particularly, for realizing the flavour dependence of the mixing angles, can be precisely determined without introducing systematic errors, whereas the errors caused by missing identifying the produced heavy flavors and the obscured direction of the produced quark cannot be avoided by the traditional determination at LEP-I and SLC. Since the cross sections of the relevant hadron production are not very great, the statistics precision for the proposed determination need to be estimated, thus more work should aim to reach necessary statistics precision when performing the determination at a super $Z$-factory.

To illustrate that the proposed method still works well, though the production cross sections are not great, and to see how precise the statistics for the proposal can be reached, as examples, the experimental signals for the produced $B_c$ meson and $\Xi_{cc}^{++}$ baryon at a super $Z$-factory (luminosity as high as $L=10^{36}cm^{-2}s^{-1}$) are roughly estimated as follows.

As shown in Table.\ref{tasy000}, approximately $2.69\times 10^7$ $B_c$ mesons and $3.95\times 10^6$ $\Xi_{cc}^{++}$ baryons may be produced per year at a super $Z$-factory. To suppress the systematic errors caused by the direction determination of the produced hadron when determining the effective electro-weak mixing angles by measuring the asymmetry observables of the doubly heavy hadron production, it would be better that, being of the `golden channel ones', only the events, which may be well-constructed, i.e., in the events the decay of the produced hadron is exclusive, and the following decays of the hadron decay products are also exclusive, are used for the determination. For example, in the $B_c$ production case, the decays: $B_c \to J/\Psi\, \pi^+$ (branching ratio about $2.0\times 10^{-3}$), $B_c \to J/\Psi\, \rho^+$ (branching ratio about $5.7\times 10^{-3}$), $B_c \to J/\Psi\, K^+$ (branching ratio about $1.4\times 10^{-4}$) etc\cite{bc2jpsi} and with the following decays, $J/\Psi \to l^+ l^- (l=\mu, e)$ (branching ratio about $12\%$ \cite{PDG}) and $\rho \to \pi\pi$ (with branching ratio almost $100\%$) \cite{PDG}) are the `golden channel ones', so about $2.5\times 10^4$ `golden channel events' for the processes $e^+e^-\to B_c +X \bigotimes B_c\to J/\psi \pi^+ \rm{or}\, J/\psi \rho^+ \rm{or}\, J/\psi K^+ \bigotimes J/\psi \to l^+l^- \rm{and}\, \rho\to\pi\pi$ may be collected per year at a super $Z$-factory (here the identification efficiency is assumed as $100\%$). The production events with decays such as $B_c \to \Psi(2S) \pi^+$ (branching ratio about $2.3\times 10^{-4}$) with the following decay $\Psi(2S) \to l^+ l^- (l=\mu, e)$ (branching ratio about $1.6\%$\cite{PDG}), $B_c \to B_s\, \pi^+$ (branching ratio about $4.6\%$), $B_c \to B_s\, \rho^+$ (branching ratio about $3.4\%$) and the decays $B_s\to J/\Psi \phi\; \rm{or}\, K^+K^-\, \rm{or}\, J/\Psi \pi^+ \pi^-$ followed, being the `golden channel ones', should also be considered, thus the `golden channel event' samples (ideal signals) should be sum of these two components in total. Moreover, as progress is achieved in relevant experiments and necessary Monte Carlo simulations on the events (the production with various decays) are completed, more channels may become `golden ones' for the determination, e.g., the channels, in which only one neutral particle is missed, probably may also become `golden ones', so finally the statistics precision for the determination of the heavy flavored effective electro-weak mixing angles may reach a quite good level. In the baryon $\Xi_{cc}^{++}$ production case, the golden channels may be of the decays $\Xi_{cc}^{++} \to \Lambda_c^+ K^- \pi^+ \pi^+$ (branching ratio about $10\%$\cite{xiccdecay}), and $\Lambda_c^+$ is reconstructed via $\Lambda_c^+ \to p K^- \pi^+$ (branching ratio $6.23\%$\cite{PDG}). The `total branching ratio' is $6.23\times 10^{-3}$, so about $2.4\cdot10^4$ $\Xi_{cc}^{++} \to \Lambda_c^+ K^- \pi^+ \pi^+$ `golden events' per year may be reconstructed. Again as experimental progress is achieved and Monte Calor simulations are completed, more mechanisms for $\Xi_{cc}^{++}$ production and more decays may be realized as 'golden ones', thus the statistics precision will be good too.

Overall, the proposed method to determine the effective electro-weak mixing angles ${\rm sin}^2\theta^{\rm f}_{\rm eff}$ (f=c,b; lept) by measuring the forward--backward, left--right, and left--right forward--backward asymmetries with the observables $A_{FB}$, $A_{LR}$ and $A^{FB}_{LR}$ of the doubly heavy hadron production at a super $Z$-factory has obvious advantages. As a proposal here it was proposed for finding the clues beyond SM. The proposed method thus may be an independent complement to traditional methods used to determine the effective electro-weak mixing angles ${\rm sin}^2\theta^{\rm f}_{\rm eff}$ (f = c,b; lept). Although the proposed method has a disadvantage, i.e., the production cross sections are not very great, high luminosity of the super $Z$-factory and effective detectors for experimental determination can alleviate it. In this work, only theoretical estimates were carried out; future work should therefore aim to make Monte Carlo simulations, to improve the estimates of the systematics errors and efficiency of the detectors etc. As long as there is a plan to build a super $Z$-factory in the world, it is worth studying and implementing the proposed method.

\vspace{4mm}

\noindent {\bf\Large Acknowledgments:} This work was supported in part by Nature Science Foundation of China (NSFC) under Grant No. 11275036, No. 11745006, No. 11535002, No. 11821505, No. 11675239, No. 11705045, No. 11821505, No. 11805140 and by Key Research Program of Frontier Sciences, CAS, Grant No. QYZDY-SSW-SYS006.

\end{document}